\documentclass[
 twocolumn,
 aps,
 prl
 ]{revtex4-1}

\usepackage{xcolor}%for todo

\usepackage{graphicx}% Include figure files
\usepackage{dcolumn}% Align table columns on decimal point
\usepackage{bm}% bold math
\usepackage{physics}
\usepackage[english]{babel}
\usepackage{amsmath}
\usepackage{amssymb}

\begin{document}
% ---newcommands----------
\newcommand{\Ftot}{F}
\newcommand{\LFP}{\hat{\mathcal{L}}_\mathrm{FP}(x,\lambda(\pspeed t))}
\newcommand{\TOexp}[1]{\overrightarrow{\exp}\left({#1}\right)}
\newcommand{\T}{\mathcal{T}}
\newcommand{\expvalaux}[1]{\expval{#1}^\dagger}
\newcommand{\Pw}{\mathcal{P}([x(t)], \pspeed)}
\newcommand{\Pwaux}{\mathcal{P}^\dagger([x(t)], \pspeed)}
\newcommand{\pspeed}{{v}}
\newcommand{\todo}[1]{(\textbf{\color{red}{#1}})}
\newcommand{\DDelta}{\Delta}

%instantaneous residence quantity
\newcommand{\Aif}{a_\T}
\newcommand{\Aim}{a(\T,\pspeed)}
%accumulated residence quantity
\newcommand{\Acf}{A_\T}
\newcommand{\Acm}{A(\T,\pspeed)}
%current observable
\newcommand{\Jf}{J_\T}
\newcommand{\Jm}{J(\T,\pspeed)}
%current observable of type I or II (decide by arg)
\newcommand{\Jtf}[1]{J^\mathrm{#1}_\T}
\newcommand{\Jtm}[1]{J^\mathrm{#1}(\T,\pspeed)}
%------------------------
\preprint{APS/123-QED}

\title{Thermodynamic Uncertainty Relation for Time-Dependent Driving\\~\\}% Force line breaks with \\

\author{Timur Koyuk and Udo Seifert\\~}
\affiliation{
{II.} Institut f\"ur Theoretische Physik, Universit\"at Stuttgart,
  70550 Stuttgart, Germany
}
\date{\today}

\begin{abstract}
Thermodynamic uncertainty relations yield a lower bound on entropy production in terms of the mean and fluctuations of a current. We derive their general form for systems under arbitrary time-dependent driving from arbitrary initial states and extend these relations beyond currents to state variables. The quality of the bound is discussed for various types of observables for an interacting pair of colloidal particles in a moving laser trap and for the dynamical unfolding of a small protein. Since the input for evaluating these bounds does not require specific knowledge of the system or its coupling to the time-dependent control, they should become widely applicable tools for thermodynamic inference in time-dependently driven systems.
\end{abstract}

\maketitle
\textsl{Introduction.} In a rough classification of non-equilibrium systems, one can distinguish
non-equilibrium steady states (NESSs), periodically driven systems and systems relaxing into equilibrium or a NESS from the vast class of systems that are driven in
some time-dependent way starting from an arbitrary initial state. A common
characteristic for all these classes is the fact that they inevitably lead to
entropy production, which is arguably the most characteristic feature that separates non-equilibrium from thermal equilibrium.
Without having detailed knowledge of the system,
however, it is not easy to determine quantitatively
the entropy
production associated with an experimentally explored non-equilibrium process  beyond the linear response regime.

The Harada-Sasa relation  as one prominent tool 
for such a quantitative inference
requires to measure the response of a NESS to an external perturbation~\cite{hara05}. 
It has successfully  been applied to, e.g.,  molecular motors~\cite{toya10} and living cells~\cite{fodo16a}.
Alternatively, from the measurement of currents in phase space the entropy production can be inferred provided the relevant
phase space is indeed accessible. In complex systems, this is a quite stringent requirement~\cite{land12,batt16}. Another strategy
is to exploit operationally accessible lower bounds on entropy production that do not require access to all
relevant degrees of freedom like the one based on the
temporal asymmetry of fluctuating trajectories~\cite{kawa07,blyt08,vaik09,rold10,muy13}.

For a NESS, a lower bound on entropy production that can be obtained from the observation of \textsl{any} 
current and its fluctuations has recently been established~\cite{bara15,ging16,seif19,horo20}. This
so-called thermodynamic uncertainty relation (TUR) holds for any system that,
on possibly some deeper unobserved level, obeys a time-continuous Markovian dynamics on discrete states or an overdamped Markovian dynamics on a 
continuous configuration space. As one immediate striking consequence, the efficiency of molecular motors can be bounded from above without knowledge 
of the specific chemo-mechanical cycles that drive the motor by observing the 
speed and its fluctuations when the motor runs against a controlled 
external force~\cite{piet16b,seif17,hwan18}. 

For periodically driven systems, inferring the entropy production, or at 
least an upper bound for it, is somewhat more complex.
There exist variants that either require time-symmetric driving~\cite{proe17} or need input from the time-reversed protocol~\cite{proe19}. In addition,
there are a number of more formal versions that  cannot easily be applied under experimentally
realistic conditions~\cite{bara18b,bara18c,koyu19}.
An operationally accessible version for arbitrary periodic driving has recently been
found that requires the response of the current to
a change of the driving frequency
as an additional input~\cite{koyu19a}. Finally, for systems relaxing either to equilibrium or to a NESS,
entropy production can be bounded by measuring the fluctuations of a current
and its mean value at the end of the observation time~\cite{dech17,liu19}.

In this Letter, we present the thermodynamic uncertainty relation for the
remaining  huge class of time-dependently driven systems mentioned at the very beginning. We 
will show how by measuring an observable, its fluctuations and its change 
under speeding up the driving parameter(s) a lower bound on the entropy production
can be obtained. The observable needs not to be a current; it could also
be, e.g., a binary variable characterizing the state of the system at the
final time or the integrated time spent in a subset of states.
As a paradigmatic illustration, we analyze in a numerical experiment the dynamical unfolding of a small peptide for which all relevant
parameters have been previously determined experimentally~\cite{stig11}. We show how a bound on the associated  entropy production can be extracted from the
observation of fluctuations without any further input.

The line-up of the genuine uncertainty relations just recalled
should be distinguished from related inequalities, called generalized
thermodynamic uncertainty relations that are a consequence of the
fluctuation theorem~\cite{hase19,timp19}. These GTURs typically yield weaker bounds on entropy
production than the TURs described above and they become trivial in the 
long-time limit. A pertinent issue with all these relations is to determine the current or
observable  that
leads to the best bound~\cite{pole16,ging16a,busi19,li19,fala20,mani20}.

The discovery of the TUR has inspired the derivation of similar relations
not necessarily involving overall entropy production for a variety of systems
including the role of finite observation times~\cite{piet17,horo17}, underdamped dynamics~\cite{dech18,fisc18,chun19,lee19},
ballistic transport between different terminals~\cite{bran18},
heat engines~\cite{shir16,piet17a,holu18,ekeh20}, stochastic field
theories~\cite{nigg20}, for the response to perturbing fields~\cite{dech18a}, for
observables that are even under time-reversal~\cite{maes17,nard17a,terl18},
for first-passage times~\cite{ging17,garr17} and for arbitrary driving~\cite{vanv20}.
Last but certainly not least, several works
have addressed how to generalize these concepts to the quantum realm, see, e.g.,~\cite{maci18,agar18,ptas18,bran18,carr19,guar19,caro19,pal20,frie20}.

\textsl{Main result for a current.}
We consider a system prepared in an arbitrary initial
state. This system is then driven through an
arbitrary control
$\lambda(\pspeed t)$ with speed parameter $\pspeed$
from $t=0$ to a final time $t=\T$. As a consequence,
the system exhibits a mean current
$\Jm$ and corresponding current fluctuations characterized by a diffusion coefficient $D_J(\T, \pspeed)$, both defined more precisely below.
Our first main result
relates these quantities with the mean total
entropy production rate $\sigma(\T, \pspeed)$ in the interval $\T$
through
\begin{equation}
{\left[\Jm + \DDelta \Jm\right]^2}
/{D_J(\T, \pspeed)} \le \sigma(\T, \pspeed).
\label{eq:ADSTUR}
\end{equation}
In comparison with the ordinary TUR for NESSs
\cite{bara15,ging16}, there is first
the dependence on the speed parameter $v$, and, second,
the crucial additional term $\DDelta \Jm$ with differential operator
\begin{equation}
\DDelta  \equiv \T\partial_\T  - \pspeed\partial_\pspeed
\label{eq:deltaJ}
\end{equation}
that describes the response of the current with respect to a slight change of the speed of driving $\pspeed$ as well as with respect to the observation time $\T$.
Consequently, all quantities entering the left-hand side of  eq.~\eqref{eq:ADSTUR} are physically transparent and thus provide an operationally accessible  lower bound
on entropy production. This result is valid for driven overdamped Langevin
dynamics of an arbitrary number of coupled degrees of freedom and for driven
Markovian systems on a discrete set of states~\footnote{
  See Supplemental Material at [SI] for the full
  derivations of the main results, details on the numerical case studies, and
  the generalization to multiple speed parameters, which includes
Refs.~\cite{lau07,risken,spec05}.}~\footnote{Coloured noise and memory can arise from integrating out degrees 
  of freedom from an underlying Markovian model. In such a case, our
  results will apply to the corresponding non-Markovian dynamics 
  as well.}.

\textsl{A first illustration: Moving trap.}
The role of the additional response term can be illustrated with an overdamped
particle with mobility $\mu$, which is dragged by a harmonic trap with stiffness $k$.
The system is initially prepared in equilibrium.
The center of the trap is moved from $x_0\equiv\lambda_0=0$ to $x_f\equiv\lambda_\T=v\T$ in time $t=\T$ with a constant velocity $\pspeed$ leading to a potential
\begin{equation}
	V(x,\lambda(\pspeed t)) = k[x-\lambda(\pspeed t)]^2/2
	\label{eq:pot_trap}
      \end{equation}
      with protocol $\lambda(\pspeed t)\equiv \pspeed t$.

One current of interest in this system is the time-averaged velocity $\nu_\T \equiv
[x(\T)-x(0)]/\T$, which is still a stochastic quantity.
Its mean, $\nu(\T,\pspeed)\equiv \expval{\nu_\T}$,
depends obviously on the observation time $\T$ and on the speed of the protocol $v$ which yields the response $\DDelta \nu(\T,\pspeed)$.

\begin{figure}[tbp]
  \centering
    \includegraphics[width=0.5\textwidth]{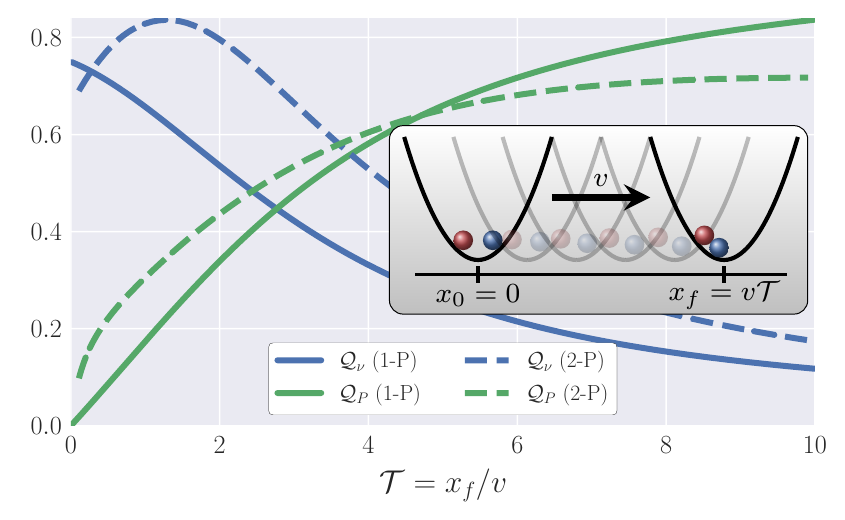}
    \caption{
        Quality factors $\mathcal{Q}_{\nu}$ and $\mathcal{Q}_P$ for velocity
        and power, respectively, as a function of inverse driving speed $\T=x_f/\pspeed$
        for a moving trap. Solid lines (1-P): One particle, $\beta=10.0$,
        $\mu=1.0$ and $x_f=\pspeed\T=10.0$. Dashed lines (2-P):
        Two interacting particles as shown in the inset. The parameters are given in
        the Supplemental Material~\cite{Note1}.
  }
  \label{fig:J_v_P}
\end{figure}

For a generic current $J$, 
the quality of bounds like ~\eqref{eq:ADSTUR} will be quantified throughout the paper by plotting
 the quality factor
\begin{equation}
	\mathcal{Q}_J\equiv\frac{\left[\Jm + \DDelta \Jm\right]^2}{D_J(\T, \pspeed) \sigma(\T, \pspeed)} \le 1.
	\label{eq:Q_J}
\end{equation}
For the particle in a moving trap, the quality factor for velocity,
$\cal Q_\nu$, is shown in Fig.~\ref{fig:J_v_P} 
as a function of
 observation time $\T$, or, equivalently, of  driving speed $\pspeed$.
The bound~\eqref{eq:ADSTUR} becomes strongest for  $\T\ll 1/(\mu k)$, i.e., for observation times
smaller than the relaxation time.
Remarkably, an estimate that yields up to $\sim 80\%$ of the total entropy production is obtained by just observing the traveled distance of the particle without knowing the strength of the trap.
In the slow-driving limit, the dispersion of the velocity becomes negligible, while heat is continuously dissipated into the surrounding medium.
As a consequence, the original TUR for a NESS is violated while  relation~\eqref{eq:ADSTUR}
holds due to the additional response term.

Another current to which relation \eqref{eq:ADSTUR} can be applied to is
the time-averaged power
\begin{equation}
	P(\T,\pspeed) = \frac{1}{\T}\int_0^\T\dd{t}\int\dd{x}
	 p(x,t;\pspeed)
        \partial_t V(x, \lambda(\pspeed t)).
\label{eq:power}
\end{equation} 
Due to the Gaussian nature of the work fluctuations, it follows that
$D_P(\T,\pspeed) = P(\T,\pspeed)/\beta$. Moreover,
the entropy production is bounded from above as
$\beta P(\T,\pspeed)/\sigma(\T,\pspeed)\geq 1$~\cite{Note1}.
Consequently, the TUR for steady-state systems 
\cite{bara15,ging16} is always violated except in the long-time limit, where the mean power converges to the mean total entropy production rate.
In contrast,
our result~\eqref{eq:ADSTUR} provides a lower bound on the mean total entropy production rate, which, in this case, is obviously
quite different from the ordinary TUR.

To illustrate the inequality~\eqref{eq:ADSTUR} for a more complex system, we
investigate two interacting particles trapped in the harmonic
potential~\eqref{eq:pot_trap}.
We choose a Lennard-Jones interaction between the particles~\cite{Note1} and
analyze the quality factors for the sum of both particle velocities, i.e., the
total traveled distance, and for the
power applied to the particles. As shown in Fig.~\ref{fig:J_v_P},
the quality factors are similar compared to
the ones for the non-interacting model and reach also about $
80\%$.

\textsl{General set-up for overdamped Langevin dynamics.}
We consider a system described by an overdamped Langevin equation for the position $x(t)$ in a thermal environment with inverse temperature $\beta$,
\begin{equation}
  \label{eq:LangevinEq}
\partial_t x(t) =  \mu \Ftot(x(t), \lambda(\pspeed t)) + \zeta(t),
\end{equation}
where $\mu$ denotes the mobility and $\zeta(t)$ is Gaussian white noise with strength $2 D\equiv 2\mu/\beta$.
The system is driven  by a force $\Ftot(x, \lambda(\pspeed t))$, which depends on an external protocol $\lambda(\pspeed t)$ that contains a speed parameter $\pspeed$.
The driving starts at $t=0$ 
with arbitrary initial distribution
 $p(x,0)$
and runs until $t=\T$. 
The time evolution of the probability density $p(x,t;v)$ follows the
Fokker-Planck equation $\partial_t p(x,t;v) = -\partial_xj(x,t;v)$
with the probability current
\begin{equation}
j(x,t;v) \equiv \left[\mu\Ftot(x,\lambda(\pspeed t))-D\partial_x\right]p(x,t;v).
\end{equation}

On the level of individual trajectories, we distinguish
\textit{state variables} from (still fluctuating)
\textit{currents}. Specifically, given a function $a(x,\lambda)$, we define
an instantaneous state variable as 
\begin{equation}
\Aif\equiv a(x(\T),\lambda(v\T)),
\label{eq:A_inst}
\end{equation}
which depends on the final value of position and control.
A further observable is its time-averaged variant given by
\begin{equation}
\Acf \equiv \frac{1}{\T}\int_0^\T\dd{t} a(x(t),\lambda(vt)).
\label{eq:residence_quantity}
\end{equation}
The ensemble average of these stochastic quantities will be denoted by
$\Aim \equiv \expval{\Aif}$ and $\Acm\equiv\expval{\Acf}$,
where we make the dependence on the two crucial parameters explicit.

For time-dependently driven systems there exist two kinds of 
currents. Both are odd under time-reversal.
The first type of current is called a \textit{jump current} and is of the form
\begin{equation}
\Jtf{I} 
= \frac{1}{\T}\int_0^\T\dd{t} d^\mathrm{I}(x(t),\lambda(\pspeed t))\circ \dot{x}(t).
\label{eq:jump_current}
\end{equation}
Here, $\circ$ denotes the Stratonovich product.
The second type is a \textit{state} current given by
\begin{equation}
  \label{eq:res_current_fluct}
\Jtf{II} 
= \frac{1}{\T}\int_0^\T\dd{t} d^\mathrm{II}(x(t),\lambda(\pspeed t)).
\end{equation}
For jump currents,
$d^\mathrm{I}(x(t),\lambda(\pspeed t))$ is an arbitrary increment, whereas for
state currents
\begin{equation}
d^\mathrm{II}(x(t),\lambda(v t)) \equiv \partial_t\lambda(\pspeed t) \partial_\lambda b(x(t),\lambda)\vert_{\lambda=\lambda(\pspeed t)}
\end{equation}
involves the derivative of a state function $b(x,\lambda$) with respect to the time-dependent driving.
We denote the mean values of these  observables by $\Jtm{I}\equiv
\expval{\Jtf{I}}$ and $\Jtm{II}\equiv \expval{\Jtf{II}}$.
A prominent example for the first type is
the mean rate of entropy production in the medium~\cite{seif12}
\begin{equation}
\sigma_m(\T, \pspeed) \equiv \frac{1}{\T}\int_0^\T\dd{t}\int\dd{x} \beta\Ftot(x,\lambda(\pspeed t)) j(x,t;\pspeed)
\end{equation}
with increment $d^\mathrm{I}(x,\lambda)=\beta\Ftot(x,\lambda)$. 
The mean total entropy production rate
\begin{equation}
  \label{eq:sigma_tot}
  \sigma(\T,\pspeed) \equiv \frac{1}{\T}\int_0^\T\dd{t}\int\dd{x}\frac{j^2(x,t;\pspeed)}{Dp(x,t;\pspeed)}
\end{equation}
additionally contains the entropy production rate of the system~\cite{seif12}.
The power applied to a system as given in eq.~\eqref{eq:power} belongs to the second type of currents and is obtained by
choosing $b(x,\lambda)=V(x,\lambda)$, where $V(x,\lambda)$ is an external
potential.

Fluctuations of all these observables can be quantified by the effective \textit{diffusion coefficient}
\begin{equation}
  \label{eq:diffusion_coefficient}
D_X(\T,\pspeed) \equiv \T\left(\expval{X_\T^2} - \expval{X_\T}^2\right)/2
\end{equation}
 and $X_\T\in\{\Aif, \Acf, \Jtf{I,II}\}$.
For both types of current observables as defined in eqs.~\eqref{eq:jump_current} and~\eqref{eq:res_current_fluct}, the TUR \eqref{eq:ADSTUR} holds true~\cite{Note1}.

\textsl{Uncertainty relation for state variables.}
Our second main result is a thermodynamic uncertainty relation for end-point and time-integrated state observables as defined in eqs.~\eqref{eq:A_inst}
and~\eqref{eq:residence_quantity}. For both types of observables, it reads~\cite{Note1}
\begin{equation}
  \label{eq:ADSTUR_instant_quant}
  [\DDelta \mathcal{A} (\T, \pspeed)]^2/D_{\mathcal{A}}(\T,\pspeed) \le \sigma(\T, \pspeed),
\end{equation}
where $\mathcal{A}(\T,\pspeed)\in \{\Aim, \Acm\}$.
For $\Aim$, 
this relation shows that a lower bound  for the mean total entropy production rate can be obtained
by just observing the final state of the system.
There is neither information required about the initial distribution
nor information about the forces acting on the particle.
This bound is especially useful for finite-time or relaxation processes
where the total entropy production is not necessarily
time-extensive.

\textsl{Sketch of the proof.}
To sketch the derivation of our main results~\eqref{eq:ADSTUR}
and~\eqref{eq:ADSTUR_instant_quant} (see \cite{Note1} for a full proof), we use a recently obtained inequality,
called the \textit{fluctuation-response inequality} (FRI),
which relates the fluctuations of an observable with its response to an external
perturbation~\cite{dech18a}.
Specifically, for this perturbation we choose the additional force
$\epsilon Y(x,t;\epsilon)$ with a parameter $\epsilon$.
Averages in the perturbed dynamics are denoted by $\expval{\cdot}^\dagger$.
For a small force, i.e., for $\epsilon\to 0$, the FRI bounds the diffusion
coefficient~\eqref{eq:diffusion_coefficient} for each choice of $X_\T$ as~\cite{dech18a,Note1}
\begin{equation}  
  \label{eq:FRI}
  D_X(\T,\pspeed) \ge \frac{\left(\partial_\epsilon
      \expval{X_\T}^\dagger\vert_{\epsilon=0}\right)^2}{1/\T\int_0^\T\dd{t}\expval{Y(x(t),
      t;\epsilon)^2/D}^\dagger\vert_{\epsilon=0}}.
\end{equation}
We choose $Y(x,t;\epsilon) =
j(x,t';v^\dagger)/p(x,t';v^\dagger)$, scale time $t'=(1+\epsilon)t$ as in
Refs.~\cite{dech18,liu19}, and additionally modify the speed parameter
$\pspeed^\dagger=\pspeed/(1+\epsilon)$.
The perturbed dynamics then corresponds to a system that evolves slightly slower or faster in time.
The denominator in~\eqref{eq:FRI} becomes the total entropy
production rate $\sigma(\T,\pspeed)$.
The nominator simplifies to $\DDelta
\mathcal{A}(\T,\pspeed)$ for state variables and to $\Jtm{I,II} + \DDelta
\Jtm{I,II} $ for currents leading to our main results~\eqref{eq:ADSTUR}
and~\eqref{eq:ADSTUR_instant_quant}.

\textsl{Generalization to discrete states: protein folding.}
Our two main results~\eqref{eq:ADSTUR} and~\eqref{eq:ADSTUR_instant_quant}
hold not only for overdamped
Langevin systems but also for systems with discrete states.
A paradigm for such a system is a protein undergoing
conformational transitions.
Experimental studies aim to infer the structure of the underlying
Markovian network that possibly contains hidden folded states.
For the protein Calmodulin, 
 the transition rates between various folded and
unfolded states  have been measured
 as a function of an external force generated by optical tweezers in Ref.~\cite{stig11}.

\begin{figure}[tbp]
  \centering
  \includegraphics[width=0.5\textwidth]{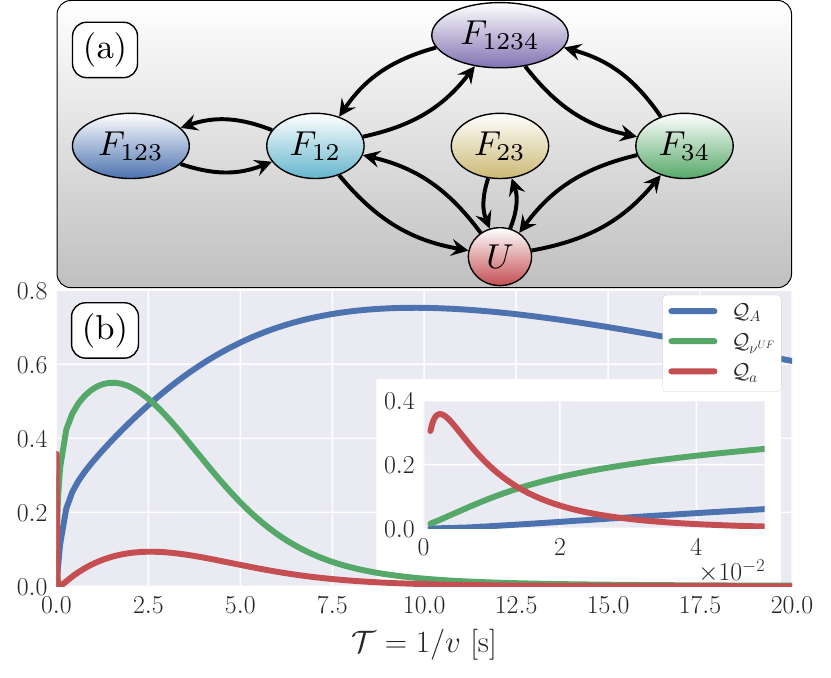}
  \caption{
  	Dynamical unfolding of Calmodulin. (a) Network of its six states, comprising an unfolded state $U$, two partially folded states $F_{12}$ and $F_{34}$, a folded state $F_{1234}$, and two misfolded states $F_{23}$ and $F_{123}$. 
  	 The force-dependent transition rates between the
  	six states as extracted from Ref.~\cite{stig11} are given in the
  	supplemental material~\cite{Note1}. (b)
  	Three quality factors as defined in the main text as a function of the inverse driving speed
        $\T=1/\pspeed$. Inset shows data for fast driving.
  }
  \label{fig:protein_folding_model}
\end{figure}

We apply our bounds to this system by using these
experimental data. In
Fig.~\ref{fig:protein_folding_model}a, the
topology of the network consisting of six different conformational
states (denoted as in the original paper) is shown. Starting in equilibrium at a
constant external force of $f_0=9.0$ pN, we drive the system in a force ramp according to 
the driving protocol
$\lambda(\pspeed t) \equiv f_0 + \pspeed t (f_1 - f_0)$ with $f_1=11.0\,\mathrm{pN}$
and
$\pspeed \T = 1.0$.

For three different observables, we consider the quality factor
of the resulting bound on the entropy production associated with this
dynamical unfolding. One estimate according to eq.~\eqref{eq:ADSTUR} is obtained by observing the
current between the unfolded state~$U$ and any of the adjacent states~$F\in\{F_{12}, F_{23},
F_{34}\}$,
\begin{equation}
  \label{eq:inference_stat_current_12}
  \nu^{UF}_{\T} \equiv
  [m_{UF}(\T)-m_{FU}(\T)]/\T.
\end{equation}
The variable $m_{UF}$ counts the total number of transitions from
the unfolded state~$U$ to any of these states~$F$ and $m_{FU}$ is the number of reverse transitions.
Two further bounds are obtained using $a(i,\lambda)=\delta_{i,F_{12}}$ in
eq.~\eqref{eq:A_inst} and $a(i,\lambda)=\delta_{i,U}$ in eq.~\eqref{eq:residence_quantity},
which corresponds to the 
characteristic function of state $F_{12}$ and $U$, respectively~\footnote{Note that $a(i,\lambda)$ is a
  straightforward generalization of the state variable $a(x,\lambda)$ defined
  in eq.~\eqref{eq:A_inst} to systems with discrete degrees of freedom, where the continuous state $x$ is replaced by the
  discrete state $i$ and the integral $\int\dd{x}$ becomes a sum
  $\sum_i$.}.
The first choice corresponds to the probability for the protein to be in state
$F_{12}$ at the end of the observation time and the latter one to the overall fraction of time 
the system has spent in the unfolded state $U$. We denote the corresponding
quality factors by $\mathcal{Q}_{a}$ and $\mathcal{Q}_{A}$, respectively.
The quality factors obtained from monitoring the mean, the
fluctuations and the response of these three observables 
are shown in Fig.~\ref{fig:protein_folding_model}b.
The quality factor $\mathcal{Q}_{A}$ becomes best at slower driving,
$\T\simeq 10$, where it yields about $75\%$ of the total
entropy production rate.
The estimate $\mathcal{Q}_{\nu^{UF}}$ through the current observable is especially strong for
intermediate times $1.0 \le \T \le 2.0$.
The quality factor $\mathcal{Q}_{a}$ based on
the observation of the final state 
is always weaker than the other two except for fast driving speeds $\T =
1/\pspeed \sim 10^{-2}$, where it reaches a maximal value of about $40\%$ as shown in the inset of
Fig.~\ref{fig:protein_folding_model}b.
Obviously, in future experiments, one should explore the bounds resulting from as many experimentally accessible state and current observables as possible since we do not yet have a criterion for selecting \textsl{a priori}  the observable
that will yield the strongest bound.

\textsl{Concluding perspective.}
We have derived a universal thermodynamic uncertainty
relation that holds for current and state variables
in systems that are time-dependently driven from an arbitrary initial state over a finite time-interval. The mean and fluctuations of
any such observable yields a lower
bound on the overall entropy production. Depending on
the conditions the observables leading to the relative best bound may change. For observables based on currents, our relation becomes the established ones
for the very special cases of time-independent driving,
of periodic driving and of relaxation at constant control parameters as
summarized in table~\ref{tab:unifying_tur}.
\begin{table}[tbp]
  \caption{Unification of TURs with their range of applicability (y=yes,
    n=no). The factor $\DDelta J(\T,\pspeed)$ in eq.~\eqref{eq:ADSTUR} 
  	specialized to  NESSs, periodic steady states (PSS) and relaxation (REL) towards equilibrium
  	or a NESS leads to
  	the terms shown on the right-hand sides in the first column.  Beyond
        these known cases, the new relation \eqref{eq:ADSTUR} is applicable
        for relaxation towards a PSS and for
  	arbitrary time-dependent driving (TTD).
      }
  \label{tab:unifying_tur}
  \centering
  \begin{tabular}{l c c c c c}
    \toprule
    $D_J\sigma/J^2\ge \Xi$ & Ref. & NESS & PSS & REL & TTD
    \\
    \colrule
    $\Xi = 1$ &\cite{bara15,ging16} & y & n & n & n \\
    $\Xi = [1 + \T \partial_\T J/J]^2$ &\cite{dech17,liu19}& y & n & y\footnote{only
                                              valid for
                                              time-independent
                                              driving
} 
& n\\
    $\Xi = [1-\Omega\partial_\Omega J/J]^2$ &\cite{koyu19a}  & y & y & n & n\\
    $\Xi = [1+\DDelta J/J]^2$ &eq.\eqref{eq:ADSTUR} & y & y & y & y\\
    \botrule
  \end{tabular}
\end{table}
In this sense, our work presents a unifying perspective on 
extant TURs.

With these relations we have provided
universally applicable tools that will allow thermodynamic inference in time-dependently driven systems. We emphasize that it is neither necessary to
know the precise coupling between the system and 
the
control nor to know the interactions within the system. It suffices that the experimentalist can change the overall speed of the control slightly and measure the resulting response of an observable. These rather weak
demands should facilitate the application to
systems beyond colloidal particles and single molecules
manipulated with 
time-dependent optical traps. Finally, as a challenge to theory, it will be
intriguing to explore whether and how these relations can be extended to
time-dependently driven  open quantum systems.

\bibliography{../Bibliography/refs.bib}
\end{document}

\preprint{APS/123-QED}
\title{Supplemental Material for "Thermodynamic uncertainty relation for time-dependent driving"\\~\\}
\author{Timur Koyuk and Udo Seifert\\~
}
\affiliation{
{II.} Institut f\"ur Theoretische Physik, Universit\"at Stuttgart,
  70550 Stuttgart, Germany
}
\date{\today}
\maketitle

This supplemental material contains five sections.
In Sect.~I, we derive the main results for current-like observables (eq.~(1)
in the main text) and for state variables (eq.~(16) in the main text) in
systems with continuous degrees of freedom.
Sect.~II contains the derivation of these main results for systems with
discrete degrees of freedom.
Sect.~IIIA and~IIIB provide more details for the case study of the moving trap
with one particle and two interacting particles, respectively.
In Sect.~IV, the rates for the protein
folding system are given. In Sect.~V, a paradigmatic example of a two-level system
illustrates how our main results can be applied to systems with multiple speed parameters.

\section{I. Derivation of the main results I: continuous degrees of freedom \label{sec:derivation_continuous}}

\subsection{Setup\label{sec:derivation_continuous_setup}}
We derive the main results for time-dependently driven systems with $\Nx$ interacting continuous
degrees of freedom.
The coordinate vector $\x\equiv\{x_1,...,x_{\Nx}\}$ describes the state of the
system, which is driven by multiple protocols
$\protocol\equiv\protocol\left(\{\pspeed_\alpha\}\right)\equiv\{\lambda_1(\pspeed_1
t),...,\lambda_{\Nlambda}(\pspeed_{\Nlambda}t)\}$ with
$\Nlambda$ speed parameter $\pspeed_\alpha$ and $\alpha\in[1,\Nlambda]$.
The dynamics obeys the Langevin equation with multiplicative noise
\begin{equation}
  \label{eq:N_dim_langevin}
  \dot{\x}(t) = \drift[\x(t),\protocol] + \stratdrift[\x(t),\protocol] +
  \sqrt{2}\Gdiffusion[\x(t),\protocol]\stratprod\noise.
\end{equation}
Here, we use the Stratonovich convention, where $\stratprod$ denotes the
Stratonovich product, $(\cdot)^T$ denotes transposition,
$\nabla\equiv\{\partial_{x_1},...,\partial_{x_{\Nx}}\}$ is the Nabla operator and $\noise \equiv
\{\noisecomponent{1},...,\noisecomponent{\Nx}\}$ is a
Gaussian white noise vector describing the random forces with mean and correlations
\begin{align}
  \label{eq:mean_noise}
  \expval{\noisecomponent{i}} &= 0,\\
  \label{eq:correlations_noise}  
  \expval{\noisecomponent{i}\noisecomponent[t']{j} } &= \delta_{ij}\delta
  (t-t').
\end{align}
Furthermore, using the $\Nx \times
\Nx$ matrix $\Gdiffusion$, we define the state- and time-dependent symmetric diffusion
matrix by $\diffusion \equiv \Gdiffusion\GdiffusionT$.
The diffusion matrix obeys the Einstein relation, i.e., $\diffusion =
\mobility/\beta$, where $\mobility$ denotes the $\Nx \times \Nx$ mobility
matrix and $\beta$ is the inverse temperature of the heat bath.
The drift vector
\begin{equation}
  \label{eq:drift_term_forces}
  \drift = \mobility \Force \equiv \mobility (-\nabla \potential + \force)
\end{equation}
contains the forces $\Force$ driving the system out of equilibrium.
They consist of a conservative force generated by a potential $\potential$ and a
non-conservative force $\force$.
Both, the drift and diffusion term are controlled by the protocol $\protocol$.
The additional drift term $\stratdrift$ arises due to the Stratonovich
convention and makes sure that a non-driven system evolves to the Boltzmann
distribution for $t\to\infty$~\cite{lau07}.

Equivalently to the Langevin equation~\eqref{eq:N_dim_langevin}, we can
describe the dynamics for the probability density $\density$ by the
Fokker-Planck equation~\cite{risken}
\begin{equation}
  \label{eq:fokker_planck_eq}
  \partial_t\density = -\nabla\current
\end{equation}
with probability current vector
\begin{equation}
  \label{eq:fokker_planck_current}
  \current\equiv \left(\drift-\diffusion\nabla\right)\density.
\end{equation}

The probability density for an individual trajectory $\x(t)$ of length
$\T$ is given by
the path weight
\begin{equation}
  \label{eq:path_weight_continuous}
  \pathweight \equiv
  \Norm\exp\left(-\action\right)\density[\x(0),0]
\end{equation}
with the action
\begin{equation}
  \label{eq:action}
  \action\equiv\frac{1}{4}\int_0^T\dd{t}\left(\dot{\x}(t)-\drift[\x(t),\protocol]\right)^T\diffusion[\x(t),\protocol]^{-1}\left(\dot{\x}(t)-\drift[\x(t),\protocol]\right)
  + \frac{1}{2}\int_0^\T\dd{t}\nabla\drift[\x(t),\protocol] +
  \Gamma[\Gdiffusion[\x(t), \protocol]]
\end{equation}
for an arbitrary initial condition $\density[\x(0),0]$.
Here, $\diffusion[\x(t),\protocol]^{-1}$ denotes the inverse of the diffusion
matrix and
\begin{equation}
  \label{eq:normalization_path_weight}
  \Norm\equiv\left(\prod_{l=1}^{\Nt}\frac{1}{\sqrt{4\pi}\determinante{\Gdiffusion[\x(t_l),{\protocol[t_l]}]}}\right)
\end{equation}
is a normalization factor with respect to the measure of integration
\begin{equation}
  \label{eq:measure_of_integration}
  \intpath \equiv \prod_{l=0}^{\Nt}\int\dd{\x(t_l)},
\end{equation}
i.e., $\int\dd{[\x(t)]}\pathweight = 1$, where $\determinante{\cdot}$ denotes
the determinant of a matrix.
The path weight in
eq.~\eqref{eq:path_weight_continuous} is well defined for a suitable
discretization in time. The trajectory is sliced into $\Nt$ discrete values
$\{\x(t_0), \x(t_1),...,\x(t_{\Nt})\}$ with $t_l\equiv l\Delta
t$, where $\Delta t$ is a small enough time step and chosen such that $\Nt\Delta
t=\T$ is fulfilled.
The two last terms in eq.~\eqref{eq:action} with
\begin{align}
  \label{eq:strato_G_matrix_action_term}
  \Gamma[\Gdiffusion] \equiv \frac{1}{2}\int_0^\T\dd{t}\sum_{ijk}&
  \delx{i}\delx{j}\left[\Gdiffusionelement{ik}\Gdiffusionelement{jk}\right] -
  \delx{i}\left[\Gdiffusionelement{kj}\delx{k}\Gdiffusionelement{ij}\right]\nonumber\\
  + &\frac{1}{2}\left\{[\delx{k}\Gdiffusionelement{ij}][\delx{i}\Gdiffusionelement{kj}]
  - [\delx{i}\Gdiffusionelement{ij}][\delx{k}\Gdiffusionelement{kj}]\right\},
\end{align}
where $\Gdiffusionelement{ij}\equiv \left[\Gdiffusion\right]_{ij}$ are matrix elements,
arise from a Stratonovich
discretization scheme~\cite{lau07}.

\subsection{Bound on the diffusion coefficient\label{sec:derivation_continuous_bound}}
For deriving our main result, we use the method introduced by Dechant and
Sasa~\cite{dech17,dech18a} that bounds the scaled cumulant generating function, or short
\textit{generating function},
\begin{equation}
  \label{eq:definition_cumulant_generating_function}
  \lambda(z)\equiv \frac{1}{\T}\ln\expval{\exp(z\T X_\T)} \equiv \frac{1}{\T}\ln\intpath
  \pathweight \exp(z\T X_\T[\x(t)])
\end{equation}
for a fluctuating observable $X_\T=X_\T[\x(t)]\in\{\Aif, \Acf, \Jtf{I,II}\}$
by introducing an auxiliary path weight $\auxpathweight$ that describes an auxiliary
dynamics obeying a Langevin equation of type~\eqref{eq:N_dim_langevin}.
The first two derivatives of $\lambda(z)$ at $z=0$ yield the mean and
diffusion coefficient of $X_\T$, i.e.,
\begin{align}
  \label{eq:lambda_d1}
  \lambda'(z)\vert_{z=0} &= \expval{X_\T},\\
    \label{eq:lambda_d2}
  \lambda''(z)\vert_{z=0} &= 2D_{X}(\T,\{\pspeed_\alpha\}).
\end{align}
Writing the expectation value
in~\eqref{eq:definition_cumulant_generating_function} in terms of the
auxiliary path weight $\auxpathweight$ and using Jensen's inequality, we get
the lower bound
\begin{equation}
  \label{eq:bound_on_generating_function}
  \lambda(z) = \frac{1}{\T}\ln\intpath \auxpathweight \frac{\pathweight}{\auxpathweight}\exp(z\T
  X_\T[\x(t)]) \ge z\expvalaux{X_\T} - \frac{1}{\T}\expvalaux{\ln\left(\frac{\auxpathweight}{\pathweight}\right)}
\end{equation}
on the generating function, where $\expvalaux{\cdot}$ denotes the expectation
value in the auxiliary dynamics.
For a suitable choice of the path weight $\auxpathweight$, the
bound~\eqref{eq:bound_on_generating_function} implies a bound on the diffusion
coefficient as we will show below.

We require the auxiliary path weight $\auxpathweight$ to follow the
Fokker-Planck equation
\begin{equation}
  \label{eq:auxiliary_fokker_planck}
  \partial_t\auxdensity = -\nabla \auxcurrent
\end{equation}
with auxiliary density $\auxdensity$ and auxiliary current
\begin{equation}
  \label{eq:auxiliary_current}
  \auxcurrent\equiv\left(\auxdrift-\auxdiffusion\nabla\right)\auxdensity.
\end{equation}
Here, the auxiliary diffusion process is generated by the drift vector
$\auxdrift$ and diffusion matrix $\auxdiffusion\equiv\auxGdiffusion[\x,\auxprotocol] [\auxGdiffusion]^T$, where we introduce the
auxiliary protocol $\auxprotocol\equiv \auxprotocol\left(\{\auxpspeed_\alpha\}\right)\equiv\{\lambda^\dagger_1(\auxpspeed_1
t),...,\lambda^\dagger_1(\auxpspeed_\Nlambda t)\}$ with auxiliary speed parameter
$\auxpspeed_\alpha$ and $\alpha\in[1,\Nlambda]$.

We choose the auxiliary dynamics such that it describes an original
dynamics that evolves slower or faster in time, i.e.,
$t\to (1+\epsilon) t $ and $\pspeed_\alpha\to \pspeed_\alpha/(1+\epsilon)$ and
that is driven with the same protocol functions $\{\lambda_\alpha\}$.
Here, $\epsilon=\order{z}$ is assumed to be a small parameter, i.e., the
auxiliary dynamics is considered in a linear response regime around
the original dynamics~\cite{dech18a}.
Hence, the protocol and the speed parameter of the auxiliary dynamics read
\begin{equation}
  \label{eq:auxiliary_protocol}
  \auxprotocol[t] = \protocol[(1+\epsilon)t]\left(\{\auxpspeed_\alpha\}\right)= \{\lambda_1(\pspeed_1
  t),...,\lambda_\Nlambda(\pspeed_\Nlambda t)\} = \protocol  
\end{equation}
and
\begin{align}
  \label{eq:auxiliary_speed_parameter}
  \auxpspeed_\alpha \equiv \pspeed_\alpha/(1+\epsilon),
\end{align}
respectively. Thus, the protocol is the same as for the original dynamics.
The auxiliary density and current are consequently given by
\begin{equation}
  \label{eq:auxiliary_density}
  \auxdensity \equiv \density[ {\x, [1+\epsilon] t;\{\auxpspeed_\alpha\}} ],
\end{equation}
and
\begin{align}
  \label{eq:auxiliary_current}
  \auxcurrent \equiv (1+\epsilon)\current[ {\x, [1+\epsilon] t;\{\auxpspeed_\alpha\}}],
\end{align}
respectively, where we assume that both processes start in the same initial condition
$\density[\x(0),0]$. Furthermore, we assume that the initial condition does
not depend on the speed parameter $\pspeed_\alpha$. If this was not the case,
like, e.g., for a system in a periodic steady-state, 
an additional boundary term would occur for finite observation times (see \cite{koyu19a}).
The auxiliary dynamics introduced above describing a ``time-scaled'' diffusion
process can be considered as a process generated by
an additional drift vector and the original diffusion matrix according to
\begin{equation}
  \label{eq:aux_drift}
  \auxdrift = \drift + \epsilon \LRDrift  
\end{equation}
and
\begin{align}
  \label{eq:aux_diffusion}
  \auxdiffusion = \diffusion, \;\auxGdiffusion = \Gdiffusion,
\end{align}
respectively.
The drift term $\auxdrift$ contains the small additional force
\begin{equation}
  \label{eq:aux_force}
  \epsilon\LRDrift\equiv \epsilon\current[{\x,
  [1+\epsilon]t;\{\auxpspeed_\alpha\}}]/\density[{\x,
  [1+\epsilon]t;\{\auxpspeed_\alpha\}}]
\end{equation}
and is called a \textit{virtual perturbation}~\cite{dech18a}.
The diffusion matrix $\diffusion$ is the same as for the original process. Thus, the
normalization constant~\eqref{eq:normalization_path_weight} is the same for
both dynamics.

Next, we insert the auxiliary path weight
\begin{equation}
  \label{eq:auxiliary_path_weight}
  \auxpathweight\equiv \Norm\exp\left(-\auxaction\right)
\end{equation}
with action
\begin{equation}
  \label{eq:auxiliary_action}
  \auxaction\equiv\frac{1}{4}\int_0^T\dd{t}\left(\dot{\x}(t)-\auxdrift[\x(t),\protocol,t]\right)^T\diffusion[\x(t),\protocol]^{-1}\left(\dot{\x}(t)-\auxdrift[\x(t),\protocol,t]\right)
  + \frac{1}{2}\int_0^\T\dd{t}\nabla\auxdrift[\x(t),\protocol,t] +
  \Gamma[\Gdiffusion[\x(t), \protocol]],
\end{equation}
and the auxiliary drift term defined in eq.~\eqref{eq:aux_drift} into
eq.~\eqref{eq:bound_on_generating_function} and obtain the bound
\begin{equation}
  \label{eq:bound_generating_function_sigma}
  \lambda(z)\ge z\expvalaux{X_\T} - \frac{\epsilon^2}{4}\sigma(\Taux,\{\auxpspeed_\alpha\}).
\end{equation}
The total entropy production rate
\begin{equation}
  \label{eq:entropy_production_rate}
  \sigma(\Taux,\{\auxpspeed_\alpha\}) \equiv
  \frac{1}{\Taux}\int_0^{\Taux}\dd{t'}\current[\x,t';\{\auxpspeed_\alpha\}]^T\diffusion[\x,{\protocol[t'](\{\auxpspeed_\alpha\})}]^{-1}\current[\x,t';\{\auxpspeed_\alpha\}]/\density[\x,
  t';\{\auxpspeed_\alpha\}]
\end{equation}
is the one of a system with observation time $\Taux \equiv (1+\epsilon) \T$ and speed parameter
$\auxpspeed=\pspeed_\alpha/(1+\epsilon)$, where we used the substitution
$t'=(1+\epsilon) t$.

We take the limit $\epsilon\to 0$ and calculate the leading orders of the
two terms in eq.~\eqref{eq:bound_generating_function_sigma}.
The first term in eq.~\eqref{eq:bound_generating_function_sigma} depends on
the observable $X_\T$ and is given by one of the following four expectation values
depending on the choice of $X_\T\in\{\Aif, \Acf, \Jtf{I,II}\} $
\begin{alignat}{2}
  {\label{eq:auxexpval_Aif}}
  \expvalaux{\Aif} &= \Aim[\Taux,\{\auxpspeed_\alpha\}] &&= \int\dd{\x}
  a(\x,\protocol[\Taux](\{\auxpspeed_\alpha\}) )
                     \density[\x,\Taux;\{\auxpspeed_\alpha\}],\\
  \label{eq:auxexpval_Acf}
  \expvalaux{\Acf} &= \Acm[\Taux,\{\auxpspeed_\alpha\}]&&=
                     \frac{1}{\Taux}\int_0^{\Taux}\dd{t'}\int\dd{\x}
                     a(\x,\protocol[t'](\{\auxpspeed_\alpha\}))\density[\x,t';\{\auxpspeed_\alpha\}],\\
  \label{eq:auxexpval_Jtf_1}
  \expvalaux{\Jtf{I}} &= \Jtm[\Taux,\{\auxpspeed_\alpha\}]{I}&&=
                     \frac{1}{\Taux}\int_0^{\Taux}\dd{t'}\int\dd{\x}
                        (1+\epsilon)\vb{d}^\mathrm{I}(\x,\protocol[t'](\{\auxpspeed_\alpha\}))\cdot\current[\x,t';\{\auxpspeed_\alpha\}],\\
  \label{eq:auxexpval_Jtf_2}
  \expvalaux{\Jtf{II}} &= \Jtm[\Taux,\{\auxpspeed_\alpha\}]{II}&&=
                        \frac{1}{\Taux}\int_0^{\Taux}\dd{t'}\int\dd{\x}
                        (1+\epsilon)d^\mathrm{II}(\x,\protocol[t'](\{\auxpspeed_\alpha\}))\density[\x,t';\{\auxpspeed_\alpha\}].
\end{alignat}
The increment
\begin{equation}
  \label{eq:reminder_increment_II}
  d^\mathrm{II}(\x,\protocol) \equiv \partial_t\protocol\cdot\nabla_{\protocol} b(\x,\protocol),
\end{equation}
involves the time-derivative of a state function $b(\x,\protocol)$. The
vector $\vb{d}^\mathrm{I}(\x,\protocol)$ is arbitrary.
Calculating the leading order in $\epsilon=\order{z}$ in
eq.~\eqref{eq:bound_generating_function_sigma} via
eqs.~\eqref{eq:entropy_production_rate}--\eqref{eq:auxexpval_Jtf_2} and optimizing
with respect to $\epsilon$ leads to a local quadratic bound on the generating
function that implies with eqs.~\eqref{eq:lambda_d1} and~\eqref{eq:lambda_d2}
our main results. These are the bounds on the diffusion coefficients
\begin{equation}
  \label{eq:bound_type_I}
  D_J(\T,\{\pspeed_\alpha\}) \ge \frac{[J(\T,\{\pspeed_\alpha\}) + \DDelta
    J(\T,\{\pspeed_\alpha\}) ]^2}{\sigma(\T,\{\pspeed_\alpha\})}  
\end{equation}
and
\begin{align}
  \label{eq:bound_type_II}
  D_{\mathcal{A}}(\T,\{\pspeed_\alpha\}) \ge \frac{[\DDelta
  \mathcal{A}(\T,\{\pspeed_\alpha\}) ]^2}{\sigma(\T,\{\pspeed_\alpha\})}
\end{align}
with
$J(\T,\{\pspeed_\alpha\})\in\{\Jtm[\T,\{\pspeed_\alpha\}]{I},\Jtm[\T,\{\pspeed_\alpha\}]{II}\}$ and
$\mathcal{A}(\T,\{\pspeed_\alpha\})\in\{\Aim[\T,\{\pspeed_\alpha\}],\Acm[\T,\{\pspeed_\alpha\}]\}$.
The differential operator
\begin{equation}
  \label{eq:diff_operator}
  \DDelta  \equiv \T\partial_\T-\sum_\alpha \pspeed_\alpha
  \partial_{\pspeed_\alpha}
\end{equation}
is a generalization of the operator defined in eq.~(2) in the main text
to multiple speed parameters.
These relations prove the inequalities eqs.~(1) and~(16) in the main text for
continuous degrees of freedom.

Finally, we connect the above introduced formalism to the FRI~(17) in the main
text by deriving it explicitly for a system with one speed parameter
$\pspeed$ and one degree of freedom $x$.
Instead of using the specific perturbation in eq.~\eqref{eq:aux_force}, we use an
arbitrary perturbation of the drift term as
\begin{equation}
  \label{eq:FRI_Drift_Pert}
  B^\dagger(x,\lambda_t,t) \equiv B(x,\lambda_t) + \epsilon Y(x, t; \epsilon),
\end{equation}
where the additional drift term $Y(x, t; \epsilon)$ depends on a parameter
$\epsilon$.
Inserting eq.~\eqref{eq:FRI_Drift_Pert} into
\eqref{eq:bound_on_generating_function} yields the lower bound on the cumulant
generating function
\begin{equation}
  \label{eq:FRI_SCGF_Bound}
  \lambda(z)\ge z\expvalaux{X_\T} -
  \frac{\epsilon^2}{4}\expvalaux{Y(x(t),t;\epsilon)^2/D},
\end{equation}
where $D$ denotes the diffusion constant. Next, we consider the limit
$\epsilon=\order{z}\to 0$, i.e., the perturbed dynamics is in a linear
response regime around the original dynamics.
Optimizing eq.~\eqref{eq:FRI_SCGF_Bound} with respect to $\epsilon$ and
calculating the leading order in eq.~\eqref{eq:FRI_SCGF_Bound} leads to a
bound on the diffusion coefficient
\begin{equation}
  \label{eq:LFRI}
  D_X(\T,\pspeed) \ge
  \frac{(\partial_\epsilon\expvalaux{X_\T}\vert_{\epsilon=0})^2}{1/\T\int_0^\T\dd{t}\expvalaux{Y(x(t),t;\epsilon)^2/D}\vert_{\epsilon=0}
  }.
\end{equation}
This inequality is called a \textit{fluctuation-response
  inequality} (FRI)~\cite{dech18a} and is identical to eq.~(17) in the main text.
Using the special choice of the force in eq.~\eqref{eq:aux_force} and
inserting eqs.~\eqref{eq:auxexpval_Aif}--\eqref{eq:auxexpval_Jtf_2} into the
FRI~\eqref{eq:LFRI} leads to our main results eqs.~\eqref{eq:bound_type_I}
and~\eqref{eq:bound_type_II}, i.e., eqs.~(1) and~(16) in the main text.
We note that the FRI in eq.~\eqref{eq:LFRI} is a special case of a FRI due to
the linear response assumption. For arbitrary strong perturbations a more
general FRI can be derived, which involves higher-order cumulants
(see Ref.~\cite{dech18a} for details).

\section{II. Derivation of the main result: discrete states}
\subsection{Setup}
Here, we consider systems with $\Nstates$ discrete states.
These systems obey a Markovian dynamics described by the master equation
\begin{equation}
  \label{eq:master_equation}
  \partial_t \probability{i} = - \sum_j \probabilitycurrent{ij}
\end{equation}
with probability current
\begin{equation}
  \label{eq:probability_current}
  \probabilitycurrent{ij}\equiv \probability{i}\rate{ij} - \probability{j}\rate{ji}.
\end{equation}
Here, $\rate{ij}$ denotes the transition rate from state $i$ to state $j$ at
time $t$, when the system is driven by the protocol $\protocol\equiv\protocol\left(\{\pspeed_\alpha\}\right)\equiv\{\lambda_1(\pspeed_1
t),...,\lambda_{\Nlambda}(\pspeed_{\Nlambda}t)\}$ with
$\Nlambda$ speed parameter $\pspeed_\alpha$ and $\alpha\in[1,\Nlambda]$.
For thermodynamic consistency, the rates
$\rate{ij}$ must fulfill the so-called \textit{local detailed balance
  condition}~\cite{seif17}
\begin{equation}
  \label{eq:local_detailed_balance_condition}
  \frac{\rate{ij}}{\rate{ji}} = \exp\left[-\beta\denergy{ij} + \drivingaffinity{ij}\right],
\end{equation}
where $\beta$ is the inverse temperature of the heat bath,
$\denergy{ij}\equiv\energy{j}-\energy{i}$ is the energy difference between
state $i$ and $j$, and $\drivingaffinity{ij}$ is a driving affinity, e.g, a
non-conservative force, which drives the system in addition to the
time-dependent energies.
Both, the energies and the driving affinities depend on the protocol
$\protocol$.

Similarly to systems with continuous degrees of freedom, we can define a
probability density for a discrete trajectory $\n$ of length $\T$ with initial
condition $\probability[0]{\n[0]}$
\begin{equation}
  \label{eq:discete_path_weigth}
  \pathweight[\n]\equiv \exp\left(-\int_0^\T\dd{t} \sum_i \exitrate{i}\occupationvariable{i}
    + \int_0^\T\dd{t}\sum_{ij}\ln\left[\rate{ij}\right]\djumpvariable{ij}\right)\probability[0]{\n[0]},
\end{equation}
where $\exitrate{i}\equiv\sum_j\rate{ij}$ is the escape, or exit rate, of state
$i$, $\occupationvariable{i}$ is a variable that is 1 if state $i$ is occupied and 0,
otherwise, and $\jumpvariable[\T]{ij}$ counts the total number of transitions
from state $i$ to state $j$.
The time derivative of the latter one
\begin{equation}
  \label{eq:djumpvariable}
  \djumpvariable{ij} \equiv \sum_l \delta(t-t^{(ij)}_l)
\end{equation}
depends on the times $t^{(ij)}_l$ at which transitions from $i$ and $j$ occur.
The path probability~\eqref{eq:discete_path_weigth} is normalized, i.e.,
$\sumpath\pathweight[\n]=1$, where $\sumpath$ denotes the summation over all
paths.
The mean values of the variables $\occupationvariable{i}$ and
$\djumpvariable{ij}$ describing a trajectory are given by
\begin{alignat}{2}
  \label{eq:mean_occ}
  \expval{\occupationvariable{i}}&\equiv\sumpath\pathweight[\n]\occupationvariable{i}
  &&= \probability{i},\\
  \label{eq:mean_djump}
  \expval{\djumpvariable{ij}} &\equiv\sumpath\pathweight[\n]\djumpvariable{ij} &&=\probability{i}\rate{ij}.
\end{alignat}

For systems with discrete states, the analogues of observables $\Aif,\Acf$ and $\Jtf{I,II}$ defined in eqs.~(9)--(12) in the main
text, are given by
\begin{align}
  \label{eq:discrete_state_var}
  \Aif    &\equiv a(\n[\T],\protocol[\T]),\\
  \label{eq:discrete_time_averaged_state_var}
  \Acf    &\equiv \frac{1}{\T}\int_0^\T\dd{t} a(\n,\protocol),\\
  \label{eq:discrete_current_I_var}
  \Jtf{I} &\equiv \frac{1}{\T}\int_0^\T\dd{t}
            \sum_{ij}d^\mathrm{I}_{ij}(\protocol)\djumpvariable{ij},\\
  \label{eq:discrete_current_II_var}
  \Jtf{II} &\equiv \frac{1}{\T}\int_0^\T\dd{t} d^\mathrm{II}(\n,\protocol),
\end{align}
where $a(\n,\protocol)$ is an arbitrary state variable,
$d^\mathrm{I}_{ij}(\protocol)=-d^\mathrm{I}_{ji}(\protocol)$ are
anti-symmetric increments and
\begin{equation}
  \label{eq:discrete_increment_II}
  d^\mathrm{II}(\n,\protocol)\equiv \partial_t\protocol\cdot\nabla_{\protocol} b(\n,\protocol)
\end{equation}
can be written as a time-derivative of a state variable $b(\n,\protocol)$.
The mean values of
eqs.~\eqref{eq:discrete_state_var}-\eqref{eq:discrete_current_II_var} are
given by
\begin{alignat}{2}
  {\label{eq:mean_discrete_state_var}}
  \expval{\Aif} &\equiv \Aim[\T,\{\pspeed_\alpha\}] &&=\sum_i a(i,\protocol[\T])\probability[\T,\{\pspeed_\alpha\}]{i},\\
  \label{eq:mean_discrete_time_averaged_state_var}
  \expval{\Acf}   &\equiv \Acm[\T,\{\pspeed_\alpha\}] &&=\frac{1}{\T}\int_0^\T\dd{t} \sum_i a(i,\protocol)\probability{i},\\
  \label{eq:mean_discrete_current_I_var}
  \expval{\Jtf{I}} &\equiv \Jtm[\T,\{\pspeed_\alpha\}]{I} &&=\frac{1}{\T}\int_0^\T\dd{t}
            \sum_{i>j}d^\mathrm{I}_{ij}(\protocol)\probabilitycurrent{ij},\\
  \label{eq:mean_discrete_current_II_var}
  \expval{\Jtf{II}} &\equiv \Jtm[\T,\{\pspeed_\alpha\}]{II} &&= \frac{1}{\T}\int_0^\T\dd{t}\sum_i d^\mathrm{II}(i,\protocol)\probability{i}.
\end{alignat}

\subsection{Bound on the diffusion coefficient}
We use the same formalism as used above for systems with continuous degrees of
freedom to bound the generating
function~\eqref{eq:definition_cumulant_generating_function} for systems with
discrete states.
The generating function for an observable
$X_\T=X_\T[\n]\in\{\Aif,\Acf,\Jtf{I,II}\}$ is bounded analogously to
eq.~\eqref{eq:bound_on_generating_function} by introducing an auxiliary
process with path weight $\auxpathweight[\n]$, where the integral $\intpath$ is replaced by the summation over all paths
$\sumpath$ in eq.~\eqref{eq:bound_on_generating_function}.

We require the auxiliary path weight $\auxpathweight[\n]$ to describe a master
equation
\begin{equation}
  \label{eq:auxiliarly_master_equation}
  \partial_t \auxprobability{i} = - \sum_j \auxprobabilitycurrent{ij}
\end{equation}
with probability current
\begin{equation}
  \label{eq:auxiliary_probability_current}
  \auxprobabilitycurrent{ij}\equiv \auxprobability{i}\auxrate{ij}-\auxprobability{j}\auxrate{ji}
\end{equation}
and auxiliary transition rates $\auxrate{ij}$.
Here, we introduce the
auxiliary protocol $\auxprotocol\equiv \auxprotocol\left(\{\auxpspeed_\alpha\}\right)\equiv\{\lambda^\dagger_1(\auxpspeed_1
t),...,\lambda^\dagger_1(\auxpspeed_\Nlambda t)\}$ with auxiliary speed parameters
$\auxpspeed_\alpha$ and $\alpha\in[1,\Nlambda]$.
By inserting the auxiliary path weight $\auxpathweight[\n]$ via
definition~\eqref{eq:discete_path_weigth} with transition rates $\auxrate{ij}$ into the discrete version of
eq.~\eqref{eq:bound_on_generating_function}, we get
\begin{equation}
  \label{eq:discrete_bound_on_generating_function}
  \lambda(z)\ge z\expvalaux{X_\T} -
  \frac{1}{\T}\int_0^\T\dd{t}\sum_{ij}\left(\auxprobability{i}\auxrate{ij}\ln\left[\frac{\auxrate{ij}}{\rate{ij}}\right]
  - \auxprobability{i}\left[\auxrate{ij}-\rate{ij}\right]\right).
\end{equation}
Here, we assume that both processes, i.e., the original and the auxiliary process,
start with the same initial condition $\probability[0]{\n[0]}$, which we
require to be independent of the speed parameters $\{\pspeed_\alpha\}$ (cf. the
discussion for systems with continuous degrees of freedom given above).

We choose the rates of the auxiliary process as
\begin{equation}
  \label{eq:auxiliary_rates}
  \auxrate{ij} \equiv \rate{ij} \left(1+\epsilon\left[ 1  -
      \eta_{ij}(\epsilon,t) \delta  \right] \right),
\end{equation}
where $\epsilon = \order{z}$ is a small parameter, $\delta$ is a free
parameter that can be chosen as 1
or 0 and
\begin{equation}
  \label{eq:eta_ij}
  \eta_{ij}(\epsilon,t)\equiv
  \frac{2\probability[{[1+\epsilon]t;\{\pspeed_\alpha/[1+\epsilon]\}}]{j}\rate{ji}}{\activity[{[1+\epsilon]t;\{\pspeed_\alpha
      /[1+\epsilon]\}}]{ij}},
\end{equation}
where
\begin{equation}
  \label{eq:orgiginal_process_activity}
  \activity{ij} \equiv   \probability{i}\rate{ij}+\probability{j}\rate{ji}
\end{equation}
is the average dynamical activity at link $ij$ of the original process.
Here, we have chosen the auxiliary protocol and speed parameter according to
eqs.~\eqref{eq:auxiliary_protocol} and~\eqref{eq:auxiliary_speed_parameter}.
This choice of rates corresponds to a ``time-scaled'' process with probability
and current given by
\begin{equation}
    \label{eq:auxiliary_probability_choice}
  \auxprobability{i} \equiv\probability[ {[1+\epsilon]t;\{\pspeed_\alpha/[1+\epsilon]\}}]{i}
\end{equation}
and
\begin{align}
  \label{eq:auxiliary_probability_current_choice}
  \auxprobabilitycurrent{ij} \equiv
  (1+\epsilon)\probabilitycurrent[ {[1+\epsilon]t;\{\pspeed_\alpha
  /[1+\epsilon]\}}]{ij},
\end{align}
respectively.

The first term in eq.~\eqref{eq:discrete_bound_on_generating_function} is
given by
\begin{alignat}{2}
  {\label{eq:discrete_mean_discrete_state_var}}
  \expvalaux{\Aif} & = \Aim[\Taux,\{\auxpspeed_\alpha\}] &&=\sum_i
  a(i,\protocol[\Taux](\{\auxpspeed_\alpha\}))\probability[{\Taux;\{\auxpspeed_\alpha\}}]{i},\\
  \label{eq:discrete_mean_discrete_time_averaged_state_var}
  \expvalaux{\Acf}   &= \Acm[\Taux,\{\auxpspeed_\alpha\}]
  &&=\frac{1}{\Taux}\int_0^{\Taux}\dd{t'} \sum_i
  a(i,\protocol[t'](\{\auxpspeed_\alpha\}))\probability[{t';\{\auxpspeed_\alpha\}}]{i},\\
  \label{eq:discrete_mean_discrete_current_I_var}
  \expvalaux{\Jtf{I}} &= \Jtm[\Taux,\{\auxpspeed_\alpha\}]{I} &&=\frac{1}{\Taux}\int_0^{\Taux}\dd{t'}
            \sum_{i>j}(1+\epsilon)d^\mathrm{I}_{ij}(\protocol[t'](\{\auxpspeed_\alpha\}))\probabilitycurrent[{t';\{\auxpspeed_\alpha\}}]{ij},\\
  \label{eq:discrete_mean_discrete_current_II_var}
  \expvalaux{\Jtf{II}} &= \Jtm[\Taux,\{\auxpspeed_\alpha\}]{II} &&=
  \frac{1}{\Taux}\int_0^{\Taux}\dd{t'}\sum_i(1+\epsilon)
  d^\mathrm{II}(i,\protocol[t'](\{\auxpspeed_\alpha\}))\probability[{t';\{\auxpspeed_\alpha\}}]{i}
\end{alignat}
depending on the choice of the observable $X_\T$.
Here, we have used the substitution $t'=(1+\epsilon)t$, $\Taux = (1+\epsilon)\T$
and $\auxpspeed_\alpha = \pspeed_\alpha /(1+\epsilon)$.
The second term in eq.~\eqref{eq:discrete_bound_on_generating_function} is
given by
\begin{equation}
  \label{eq:discrete_second_term}
  -\frac{\epsilon^2}{2\T}\int_0^\T\dd{t}\sum_{ij}\frac{\left[\probability{i}\rate{ij}+\probability{j}\rate{ji}(1-2\delta)\right]^2}{\activity{ij}}
  + \order{\epsilon^3}.
\end{equation}
An expansion for small $\epsilon$ of the r.h.s of
eq.~\eqref{eq:discrete_bound_on_generating_function} and an optimization with
respect to $\epsilon$ leads to the bounds on the diffusion coefficient
\begin{equation}
    \label{eq:bound_type_I}
  D_J(\T,\{\pspeed_\alpha\}) \ge \frac{[J(\T,\{\pspeed_\alpha\}) + \DDelta
    J(\T,\{\pspeed_\alpha\}) ]^2}{C_\delta(\T,\{\pspeed_\alpha\})}
\end{equation}
and
\begin{align}
    \label{eq:bound_type_II}
  D_{\mathcal{A}}(\T,\{\pspeed_\alpha\}) \ge \frac{[\DDelta
  \mathcal{A}(\T,\{\pspeed_\alpha\}) ]^2}{C_\delta(\T,\{\pspeed_\alpha\})}
\end{align}
with the observables
$J(\T,\{\pspeed_\alpha\})\in\{\Jtm[\T,\{\pspeed_\alpha\}]{I},\Jtm[\T,\{\pspeed_\alpha\}]{II}\}$,
$\mathcal{A}(\T,\{\pspeed_\alpha\})\in\{\Aim[\T,\{\pspeed_\alpha\}],\Acm[\T,\{\pspeed_\alpha\}]\}$
and the cost term
\begin{equation}
  \label{eq:cost_term}
  C_\delta(\T,\{\pspeed_\alpha\})\equiv \frac{2}{\T}\int_0^\T\dd{t}\sum_{ij} \frac{\left[\probability{i}\rate{ij}+\probability{j}\rate{ji}(1-2\delta)\right]^2}{\activity{ij}}.
\end{equation}
For the choice $\delta = 1$, the cost term in eq.~\eqref{eq:cost_term} is equal
or smaller than the total entropy production rate
\begin{equation}
  \label{eq:total_entropy_production_rate_discrete}
  \sigma(\T,\{\pspeed_\alpha\}) \equiv \frac{1}{\T}\int_0^\T\dd{t}\sum_{i>j}\probabilitycurrent{ij}\ln\left[\frac{\probability{i}\rate{ij}}{\probability{j}\rate{ji}}\right],
\end{equation}
i.e.,
\begin{equation}
  \label{eq:sigma_inequality}
  C_{\delta=1}(\T,\{\pspeed_\alpha\})\le \sigma(\T,\{\pspeed_\alpha\}),
\end{equation}
which can be shown by using the log-mean inequality
$(\gamma_1-\gamma_2)\ln(\gamma_1/\gamma_2)\ge
2(\gamma_1-\gamma_2)^2/(\gamma_1+\gamma_2)$ for arbitrary $\gamma_1,\gamma_2 >
0$.
Combined with eqs.~\eqref{eq:bound_type_I} and~\eqref{eq:bound_type_II}, the
inequality~\eqref{eq:sigma_inequality} proves our main results eqs.~(1)
and~(16) in the main part for systems with discrete states.

\subsection{Bound on the total average dynamic activity}
Additionally, from eqs.~\eqref{eq:bound_type_I} and~\eqref{eq:bound_type_II}, we obtain bounds on the total average dynamic activity
\begin{equation}
  \label{eq:total_average_dynamical_activtiy}
  \Act \equiv\frac{1}{\T}\int_0^\T\dd{t}\sum_{i>j}\activity{ij} = \frac{1}{\T}\int_0^\T\dd{t}\sum_{i>j}\probability{i}\rate{ij}+\probability{j}\rate{ji}
\end{equation}
by choosing $\delta=0$ in eq.~\eqref{eq:cost_term}, i.e.,
$C_{\delta=0}(\T,\{\pspeed_\alpha\}) =\Act(\T,\{\pspeed_\alpha\})$.

Finally, we show that the above derived bounds on the average dynamic
activity $\Act(\T,\{\pspeed_\alpha\})$ can also be applied to jump
observables of the type
\begin{equation}
  \label{eq:jump_observable}
  \Xcf \equiv \frac{1}{\T}\int_0^\T\dd{t}\sum_{ij}d^\mathrm{III}_{ij}(\protocol)\djumpvariable{ij},
\end{equation}
where $d^\mathrm{III}_{ij}$ are arbitrary increments that do not necessarily have to be
symmetric or anti-symmetric. With $\delta=0$ in the ansatz~\eqref{eq:auxiliary_rates},
the first term in eq.~\eqref{eq:discrete_bound_on_generating_function} becomes
\begin{equation}
  \label{eq:jump_observable_discrete}
  \expvalaux{\Xcf} =
  \frac{1}{\T}\int_0^\T\dd{t}(1+\epsilon)d^\mathrm{III}_{ij}(\protocol)\probability[{[1+\epsilon]t;\{\pspeed_\alpha
    /[1+\epsilon]\}}]{i}\rate{ij}.
\end{equation}
An expansion for small $\epsilon$ of the r.h.s of
eq.~\eqref{eq:discrete_bound_on_generating_function} and an optimization with
respect to $\epsilon$ leads to
\begin{equation}
  \label{eq:bound_on_activity}
  D_\chi(\T,\{\pspeed_\alpha\}) \ge \frac{\left[\Xcm + \DDelta \Xcm\right]^2}{2\Act(\T,\{\pspeed_\alpha\})},
\end{equation}
where
\begin{equation}
  \label{eq:diffusion_coeff_jump_obs}
  D_\chi(\T,\{\pspeed_\alpha\}) \equiv \T\left(\expval{\Xcf^2}-\expval{\Xcf}^2\right)/2
\end{equation}
is the diffusion coefficient of the jump observable $\Xcf$ and 
$\Act(\T,\{\pspeed_\alpha\})$ is the total average dynamical activity
defined in eq.~\eqref{eq:total_average_dynamical_activtiy}.
The bound in eq.~\eqref{eq:bound_on_activity} is a
generalization of bounds that were obtained in~\cite{garr17} for steady-state
systems, in~\cite{terl18} for relaxation processes and in~\cite{koyu19a} for periodically
driven systems to arbitrary
time-dependently driven systems.

\section{IIIa. Moving trap: one particle}
In this section we derive expressions for the mean particle current
$\nu(\T,\pspeed)$, the mean power $P(\T,\pspeed)$, the diffusion
coefficients $D_{\nu,P}$ and the total entropy production
rate $\sigma(\T,\pspeed)$ for the moving trap model with one particle discussed in the main text.

\subsection{Fokker-Planck equation and solution}
The Fokker-Planck equation for the moving trap reads
\begin{equation}
  \label{eq:fokker_planck_moving_trap}
  \partial_t\density[x,t;\pspeed] = -\partial_x\left(-\mu k[x-\lambda(\pspeed t)] - D
    \partial_x\right)\density[x,t;\pspeed],
\end{equation}
with protocol $\lambda(vt)=vt$ and diffusion constant $D\equiv \mu/\beta$.
The solution of eq.~\eqref{eq:fokker_planck_moving_trap} is a Gaussian distribution
\begin{equation}
  \label{eq:density_solution}
  \density[x,t;\pspeed]\equiv \frac{1}{2\pi y^2_t}\exp\left(-[x-c_t]^2/[2y^2_t]\right)
\end{equation}
with mean
\begin{equation}
  \label{eq:solution_mean}
  c_t \equiv c(t;\pspeed) \equiv \expval{x(t)}  
\end{equation}
and variance
\begin{equation}
  \label{eq:solution_variance}
  y^2_t \equiv y^2(t;\pspeed) \equiv \expval{x^2(t)} - \expval{x(t)}^2.
\end{equation}
In general, both, mean and variance depend on the speed $\pspeed$.
The system is initially prepared in equilibrium, i.e., $c_0=0$ and
$y^2_{t=0}=1/(\beta k)$.
Consequently, mean and variance are given by
\begin{align}
  \label{eq:solution_mean_expression}
  c_t &= vt - \frac{v}{\mu k}\left[1-\exp\left(-\mu k t\right)\right],\\
  \label{eq:solution_variance_expression}  
  y^2_t &= 1/(\beta k).
\end{align}
With these expressions the probability current can be written as
\begin{equation}
  \label{eq:solution_probability_current}
  j(x,t;\pspeed) = v(1-\exp\left[-\mu k t\right])\density[x,t;\pspeed].
\end{equation}

\subsection{Mean values and response terms}
Using eq.~\eqref{eq:solution_mean_expression}, the mean value of the velocity of the particle is given by
\begin{equation}
  \label{eq:mean_velocity}
  \nu(\T,\pspeed) = \expval{x(\T)}/\T = c_\T/\T =
  v\left(1-\frac{1}{\mu k \T}\left[1-\exp\left(-\mu k \T\right)\right]\right).
\end{equation}
The response term becomes
\begin{equation}
  \label{eq:response_term_velocity}
  \DDelta \nu(\T,\pspeed) \equiv (\T \partial_{\T}-v\partial_v) \nu(\T,\pspeed)
  = v\left(\frac{2}{\mu k \T}\left[1-\exp\left(-\mu k
        \T\right)\right]-\exp\left[-\mu k \T\right] - 1\right).
\end{equation}
The mean value of the power reads
\begin{equation}
  \label{eq:mean_power}
  P(\T,\pspeed) = \frac{1}{\T}\int_0^\T\dd{t}\expval{-kv(x(t)-vt)} =
  \frac{v^2}{\mu}\left(1-\frac{1}{\mu k \T}\left[1-\exp\left(-\mu k \T\right)\right]\right)
\end{equation}
with the response term
\begin{equation}
  \label{eq:response_term_power}
  \DDelta P(\T,\pspeed) \equiv (\T\partial_{\T}-v\partial_v) P(\T,\pspeed) =
  -2P(\T,\pspeed) - \frac{v^2}{\mu}\exp(-\mu k \T) + \frac{v^2}{\mu^2 k\T}
  (1-\exp[-\mu k \T]).
\end{equation}
The mean total entropy production rate can be calculated by using
eq.~\eqref{eq:solution_probability_current} and is given by
\begin{equation}
  \label{eq:mean_entropy}
  \sigma(\T,\pspeed) \equiv
  \frac{1}{\T}\int_0^\T\dd{t}\int_{-\infty}^\infty\dd{x}
  \frac{j(x,t;\pspeed)^2}{D\density[x,t;\pspeed]}= \frac{v^2}{\T
    D}\left(\T - \frac{2}{\mu k} [1-\exp(-\mu k\T)] + \frac{1}{2\mu
      k}[1-\exp(-2\mu k\T)]\right).
\end{equation}

\subsection{Diffusion coefficients}
For the diffusion coefficients
\begin{equation}
  \label{eq:diffusion_coeffcients}
  D_J(\T,\pspeed) \equiv \T \left(\expval{\Jf^2}-\expval{\Jf}^2\right)/2
\end{equation}
the term $\expval{\Jf^2}$ must be calculated.
The correlation function $\expval{\dot{x}(t)\dot{x}(t')}$ that enters in the
correlation $\expval{\nu^2_\T}$ for the velocity can be written in
terms of correlations between state functions as
\begin{align}
  \label{eq:xdot_xdot_correlation}
  \expval{\dot{x}(t)\dot{x}(t')} = 2D\delta(t-t') &+ \expval{[2\nu (x(t'),t')-\mu
    F(x(t'),\lambda(vt'))]\mu F(x(t),\lambda(vt)) }\Theta(t-t')\\ &+
  \expval{[2\nu (x(t),t)-\mu
      F(x(t),\lambda(vt))]\mu F(x(t'),\lambda(vt')) }\Theta(t'-t). \nonumber
\end{align}
Hence, in both correlation functions, i.e., in $\expval{\nu^2_\T}$ for the
velocity and in $\expval{P^2_\T}$ for the power, correlation functions $\expval{x(t)x(t')}$ occur that can be directly evaluated by solving the Langevin
equation and taking the average over all noise realizations.
Inserting these expressions into eq.~\eqref{eq:diffusion_coeffcients} yields
the diffusion coefficient
\begin{equation}
  \label{eq:D_velocity}
  D_\nu(\T,\pspeed) = \frac{D}{\mu k \T}\left(1-\exp\left[-\mu k \T\right]\right)
\end{equation}
for the velocity
and
\begin{equation}
  \label{eq:D_power}
  D_P(\T,\pspeed) =
  \frac{v^2}{\beta \mu}\left(1-\frac{1}{\mu k \T}\left[1-\exp\left(-\mu k
        \T\right)\right]\right) = P(\T,\pspeed)/\beta
\end{equation}
for the power.
The latter relation between the diffusion coefficient of the power and its
mean value arises due to the Gaussian nature of the work statistics~\cite{spec05}.
With
eqs.~\eqref{eq:mean_velocity}--\eqref{eq:mean_entropy},~\eqref{eq:D_velocity}
and \eqref{eq:D_power} the quality factors $\mathcal{Q}_{P,\nu}$ defined in
eq.~(5) in the main part can be calculated.

\subsection{Violation of the TUR for steady-state systems}
In this section, we will show that the TUR for steady-state systems
\begin{equation}
  \label{eq:TUR_SS}
  \frac{D_J\sigma}{J^2} \ge 1
\end{equation}
is strictly violated for the power, i.e., $J=P(\T,\pspeed)$.
With eq.~\eqref{eq:D_power} the l.h.s of \eqref{eq:TUR_SS} can be written as
$\sigma(\T,\pspeed)/[\beta P(\T,\pspeed)]$ for the moving trap.
Thus, if the TUR for steady-state systems was valid, it would state $\sigma(\T,\pspeed) \ge \beta P(\T,\pspeed)$.
However, as we now show in the case for the moving trap, the power rather fulfills
the opposite inequality
\begin{equation}
  \label{eq:bound_power}
  \sigma(\T,\pspeed) \le \beta P(\T,\pspeed).
\end{equation}
First, the mean entropy of the system $\expval{S_\mathrm{sys}}$ is constant
in time due to the fact that the variance $y^2_t$ is constant.
Hence, the  mean entropy change of the system $\expval{\Delta S_\mathrm{sys}}$
is zero.
Second, using
\begin{equation}
  \label{eq:moving_trap_postive_internal_energy}
  \expval{\Delta V(x(t),t)} = \expval{V(x(t),t) - V(x(0),0)} =
  \frac{k}{2}\expval{x^2(t)} \ge 0
\end{equation}
the first law of thermodynamics leads to
\begin{equation}
  \label{eq:inequality_first_law}
  \sigma(\T,\pspeed) = \frac{1}{\T}\expval{\Delta S_\mathrm{tot}} =
  \frac{1}{\T}\left(\beta \expval{Q} + \expval{\Delta S_\mathrm{sys}}\right) = \beta P(\T,\pspeed) - \expval{\Delta
    V(x(t),t)}/\T \le \beta P(\T,\pspeed),
\end{equation}
where $\expval{Q}$ is the mean heat dissipated in the medium.

\section{IIIb. Moving trap: two interacting particles}
In this section, we give further details about the moving trap with two
interacting particles in a harmonic potential. 
The two particles with mobility $\mu$ are embedded
in a solution with temperature $\beta$ and dragged by a harmonic trap with
velocity $\pspeed$ up to time $\T$ as shown in the inset of Fig.~1 in the main text.
The positions $x_1(t)$ and $x_2(t)$ of the two interacting particles obey a
set of coupled Langevin equations
\begin{align}
  \dot{x}_1(t) &= \mu \left[ - \partial_{x_1}V_\mathrm{ext}(x_1, \lambda(\pspeed t))\vert_{x_1=x_1(t)} -
                 \partial_{x_1} V_\mathrm{int}(x_1, x_2(t))\vert_{x_1=x_1(t)}\right] + \zeta_1(t),  \label{eq:LJ_x1} \\
  \dot{x}_2(t) &= \mu \left[ -\partial_{x_2}V_\mathrm{ext}(x_2, \lambda(\pspeed t))\vert_{x_2=x_2(t)}-\partial_{x_2}V_\mathrm{int}(x_1(t), x_2)\vert_{x_2=x_2(t)}\right] + \zeta_2(t),  \label{eq:LJ_x2}
\end{align}
where the noises fulfill
\begin{align}
  \expval{\zeta_i(t)} &= 0,
  \label{eq:noise_12_mean}\\
  \expval{\zeta_i(t)\zeta_j(t')} &= 2D \delta(t-t')
  \label{eq:noise_12_correlation}
\end{align}
with $i,j\in\{1,2\}$ and $D=\mu/\beta$.
The external potential generated by the harmonic trap reads
\begin{equation}
  \label{eq:ext_pot}
  V_\mathrm{ext}(x, \lambda(\pspeed t)) \equiv k(x - \lambda(\pspeed t))^2/2,
\end{equation}
where $k$ denotes the stiffness of the trap and $\lambda(\pspeed t)\equiv vt$
is the protocol.
The interaction between the particles is chosen as a Lennard-Jones potential
\begin{equation}
  \label{eq:lennard_jones_potential}
  V_\mathrm{int}(x_1,x_2) = V_0 \left[\left(\frac{r_\mathrm{min}}{\abs{x_1-x_2}}\right)^{12}-2\left(\frac{r_\mathrm{min}}{\abs{x_1-x_2}}\right)^{6}\right]
\end{equation}
with $V_0 > 0$ the depth of the potential well and $r_\mathrm{min}$ the
distance at which the potential reaches its minimum $-V_0$.

For calculating the quality factors $\mathcal{Q}_P$ and $\mathcal{Q}_\nu$
shown in Fig. 1 in the main text, we perform a Langevin simulation.
As initial condition, we choose two independent Gaussian distributions
\begin{equation}
  p_1(x_1,t=0) = \frac{1}{\sqrt{2\pi y_1^2}}\exp\left(-\frac{\left[x_1 - c_1\right]^2}{2y_1^2}\right)
  \label{eq:p_0_int_x_1}  
\end{equation}
and
\begin{align}
p_2(x_2,t=0) = \frac{1}{\sqrt{2\pi y_2^2}}\exp\left(-\frac{\left[x_2 - c_2\right]^2}{2y_2^2}\right).
  \label{eq:p_0_int_x_2}
\end{align}
In the Langevin simulation, we fix the values of the means and standard deviations to $c_1 =
2.0$, $c_2=-2.0$ and $y_1 = y_2 = 0.1$, respectively.
We further keep the parameters $\mu=1.0, k=1.0, r_\mathrm{min}=1.0, V_0=1.0, \pspeed
\T = 10.0$ and $\beta=10.0$ fixed and vary the speed of driving $\pspeed$ and
the observation time $\T$ as shown in Fig.~1 in the main text.

\section{IV. Transition rates for the unfolding of Calmodulin }
In this section, we give further information about the transition rates
between two conformational mesostates of Calmodulin that we extract from the
experimental data from Ref.~\cite{stig11} as follows.
In this experiment, an external force is applied to Calmodulin through optical
tweezers.
At constant force $f$ the folding and unfolding processes are
observed and the force-dependent transition rates are measured.
The rates at zero force are extrapolated using appropriate
models for the folding and unfolding processes (see
supplemental material of~\cite{stig11}).

\begin{figure}[tbp]
 \centering
  \includegraphics[width=0.5\textwidth]{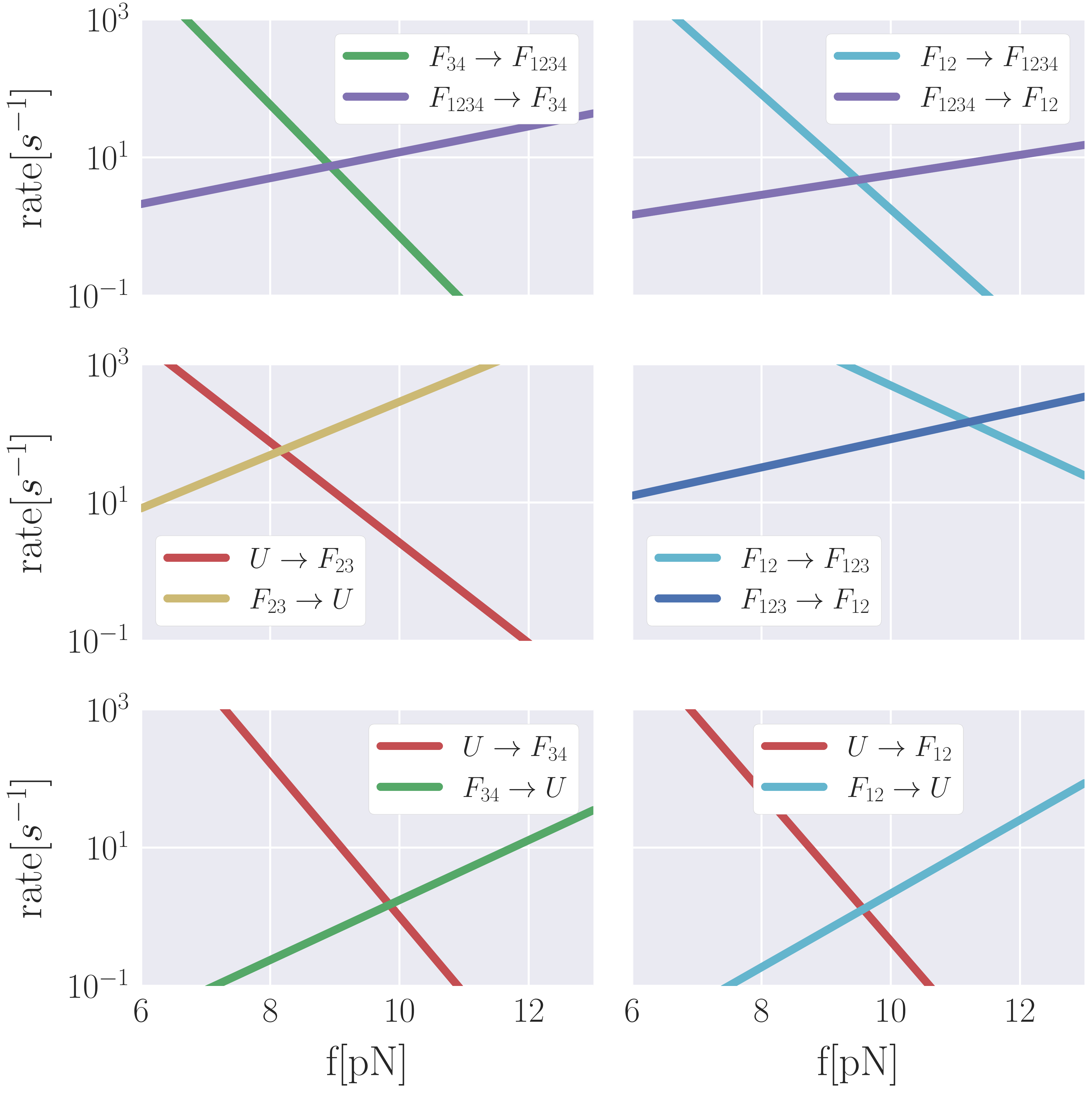}%../figures/
  \caption{Force-dependent transition rates for Calmodulin
    as used for the simulation in the main paper. Similar curves can be found in Fig.~S8 of the supplemental
    material of Ref.~\cite{stig11}.}
  \label{fig:calmodulin_rates}
\end{figure}

We parameterize the transition rates according to
\begin{equation}
  \label{eq:calmodulin_rates_fit_function}
  k_{ij}(f) \equiv k^{(0)}_{ij} \exp\left(\kappa_{ij} f\right),
\end{equation}
where $i,j\in\{U, F_{12}, F_{23},
F_{34}, F_{1234}\}$ with $i\neq j$ and $k^{(0)}_{ij}$ and $\kappa_{ij}$ are parameters listed for each
individual link in
Table~\ref{tab:fit_parameters_calmodulin}.
\begin{table}[tbp]%tbp
  \caption{Parameters for the transition rates~\eqref{eq:calmodulin_rates_fit_function} as extracted from the experimental
    data given in Ref.~\cite{stig11}.}
  \centering
  \label{tab:fit_parameters_calmodulin}
  \begin{tabular}{l c c || l c c}
    \toprule
    transition & $k^{(0)}_{ij} \,[\mathrm{s}^{-1}]$ & $\kappa_{ij}\,
                                                      [\mathrm{pN}^{-1}]$ & transition & $k^{(0)}_{ij} \,[\mathrm{s}^{-1}]$ & $\kappa_{ij}\, [\mathrm{pN}^{-1}]$\\
    \toprule
    $F_{1234} \rightarrow F_{12}$ & $0.2$ & $0.33$ & $F_{12} \rightarrow U$ & $1 \times 10^{-5}$ & $1.2$ \\
    $F_{12} \rightarrow F_{1234}$ & $3.8 \times 10^{8}$ & $-1.9$ & $U \rightarrow F_{12}$ & $2.9 \times 10^{10}$ & $-2.5$\\
    \colrule
    $F_{1234} \rightarrow F_{34}$ & $0.16$ & $0.43$ &     $F_{34} \rightarrow U$ & $7.9 \times 10^{-5}$ & $1$\\
    $F_{34} \rightarrow F_{1234}$ & $2.4 \times 10^{9}$ & $-2.2$ &     $U \rightarrow F_{34}$ & $1.3 \times 10^{11}$ & $-2.6$\\
    \colrule
    $F_{123} \rightarrow F_{12}$ & $0.74$ & $0.47$ &     $F_{23} \rightarrow U$ & $0.04$ & $0.89$\\
    $F_{12} \rightarrow F_{123}$ & $1.1 \times 10^{7}$ & $-1$ &     $U \rightarrow F_{23}$ & $4.9 \times 10^{7}$ & $-1.7$\\
    \botrule
  \end{tabular}
\end{table}
We determine the parameters by using the zero force rates listed in
Table~2 in Ref.~\cite{stig11} and Fig.~S8 in
the supplemental material of Ref.~\cite{stig11}.
The exponential force dependence in~\eqref{eq:calmodulin_rates_fit_function}
is used for the unfolding rates in Ref.~\cite{stig11}.
For the folding rates, a more complex model was used in Ref.~\cite{stig11} to
fit the experimental data.
However, in the region around $6-13\,\mathrm{pN}$ using~\eqref{eq:calmodulin_rates_fit_function} is a sufficient
approximation to reproduce the experimental measured folding rates.
The force-dependent transition rates $k_{ij}(f)$ thus obtained are shown in Fig.~\ref{fig:calmodulin_rates}.

For the time dependent driving, we vary the force $f(t) \equiv
\lambda(vt) = f_0 + \pspeed t (f_1 -f_0)$, where $\pspeed$ is the speed parameter, $f_1$ is the final force and $f_0$ is
the initial force applied to the protein.

\section{V. Generalization to multiple control parameters: two-state system}
In this section, we illustrate the generalization of our two main results
eqs.~(1) and (16)
from the main text to systems that are driven by a set of control parameters
$\{\lambda_\alpha(v_\alpha t)\}$. In this
case, the operator defined in eq.~(2) of the main text must be replaced by
eq.~\eqref{eq:diff_operator} from this Supplemental Material.

For a simple illustration, we consider a
two-state system initially prepared in equilibrium and driven through time-dependent
energy levels
\begin{align}
E_i(\boldsymbol{\lambda}_t) &\equiv E^i_0\left[1 - \exp(-\pspeed_\alpha t)\right],\label{eq:two_level_E_i}
\end{align}
with $\pspeed_{\alpha}$ the speed control parameter, where $\alpha=i$ and $E^{i}_0$ the
amplitude of driving for state $i\in\{1,2\}$.
We choose the rates between two state $i$ and $j$ as
\begin{equation}
k_{ij}(\boldsymbol{\lambda}_t) = k_0 \exp{-0.5\beta [E_j(\boldsymbol{\lambda}_t)-E_i(\boldsymbol{\lambda}_t)]},
\end{equation}
with $k_0$ as basic time-scale.
In this model, the protocol depends on two speed parameter, i.e.,
$\boldsymbol{\lambda}_t \equiv \{\lambda_1(\pspeed_1 t), \lambda_2(\pspeed_2 t)\}$.
We keep the final value of the protocol fixed, i.e.,
$\pspeed_1\T=\mathrm{const}$ and $\pspeed_2\T=\mathrm{const}$.

\begin{figure}[tbp]
  \centering
  \includegraphics[width=0.5\textwidth]{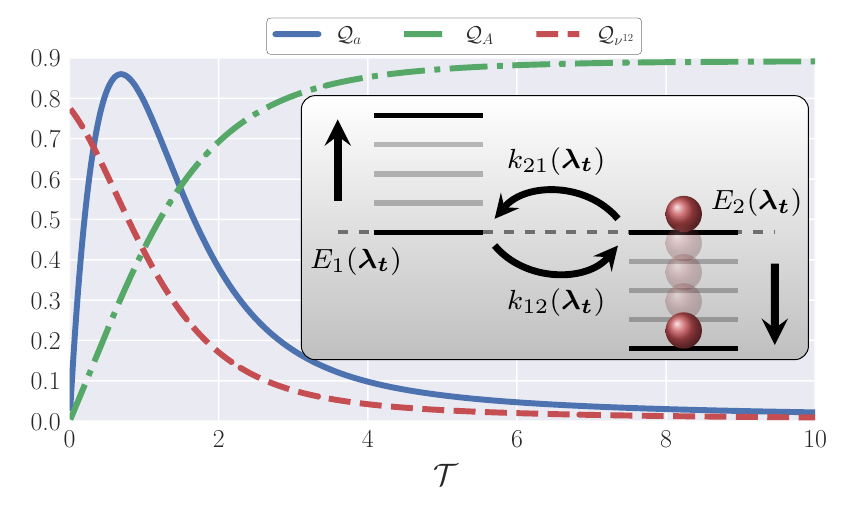}%../figures/
  \caption{
    Quality factors in the two-state system (inset) for three different classes of
    observables as a function of observation time $\T$.
    The parameters $k_0=1.0, E^1_0=-1.0, E^2_0=1.0, \beta=1.0,\pspeed_1\T=0.5$,
    and $\pspeed_2\T=1.5$ are kept fixed.}
  \label{fig:Q_two_level_3_obs}
\end{figure}

For three different observables, we consider the quality
of the resulting bound. One estimate for the total entropy
production using eq.~(1) from the main text is obtained by observing the
current between state 1 and 2
\begin{equation}
\label{eq:inference_stat_current_12}
\nu^{12}_{\T} \equiv
[m_{12}(\T)-m_{21}(\T)]/\T,
\end{equation}
where the variable $m_{ij}(\T)$ counts the total number of transitions from
state $i$ to state $j$.
Two more bounds are obtained using $a(n(t),\lambda)=\delta_{n(t),2}$ in
eqs.~\eqref{eq:discrete_state_var} and~\eqref{eq:discrete_time_averaged_state_var},
which corresponds to the 
characteristic function of state 2 either at the end of the observation time or
time-averaged. The first choice corresponds to the probability to be in state
2 and the latter one to the fraction of time 
the system spends in this state. We denote the corresponding
quality factors by $\mathcal{Q}_{a}$ and $\mathcal{Q}_{A}$, respectively.
The quality factors obtained from monitoring the mean, the
fluctuation and the response of these three observables 
are shown in  Fig.~\ref{fig:Q_two_level_3_obs}.
For fast driving $\T\ll 1$, the current observable $\nu^{12}_\T$
yields the best estimate for the total entropy production, whereas for
intermediate speeds of driving $\T\sim 1/k_0$,
the  observable based on the final state yields the best 
bound.
In the limit of quasi-static driving, the fraction
of time 
spent in state 2 yields up to 90\% of
the total entropy production rate. Throughout the whole range of driving speeds, the bounds based on these three
observables yield at least
60\% of the total entropy production rate.

\newpage
\bibliography{../../Bibliography/refs.bib} %TK 